%% file: wh_prl.tex
\documentclass[aps, prl, twocolumn, superscriptaddress, showpacs,
showkeys, amsmath, amssymb, floatfix]{revtex4} 

\usepackage{graphicx}% Include figure files 
\usepackage{dcolumn}% Align table columns on decimal point 
\usepackage{bm}% bold math 
 
\newcommand{\met}{/\!\!\!\!E_{T}} 
  
\begin{document} 

\preprint{FERMILAB-PUB-05-549-E \\ CDF/PUB/EXOTIC/CDFR/7811}
 
\title{ Search for Higgs Bosons Decaying to \boldmath$b\bar{b}$ and Produced in
  Association with \boldmath$W$ Bosons in $p\bar p$ Collisions
  at \boldmath$\sqrt{s}=1.96\,\textrm{TeV}$ }
 
\input{July_2005_Authors}

\date{\today} 
 
\begin{abstract} 
  We present a search for Higgs bosons decaying into
  $b\bar{b}$ and produced in association with $W$ bosons in $p\bar{p}$
  collisions at $\sqrt{s}=1.96\,\textrm{TeV}$. This search uses
  $320\,\textrm{pb}^{-1}$ of the dataset accumulated by the upgraded
  Collider Detector at Fermilab. Events are selected that have a
  high-transverse momentum electron or muon, missing transverse
  energy, and two jets, at least one of which is consistent with the
  hadronization of a $b$ quark.  Both the number of events and the
  dijet mass distribution are consistent with standard model
  background expectations, and we set $95\%$ confidence level upper
  limits on the production cross section times branching ratio for the
  Higgs boson or any new particle with similar decay kinematics.
  These upper limits range from $10\,\textrm{pb}$ for
  $m_H=110\,\textrm{GeV}/c^2$ to $3\,\textrm{pb}$ for
  $m_H=150\,\textrm{GeV}/c^2$.
\end{abstract} 
 

\maketitle 

A central problem in elementary particle physics is the mechanism of
electroweak symmetry breaking, by which weak vector bosons and massive
fermions acquire
non-zero masses.  Physical manifestations of electroweak symmetry
breaking may include the standard model (SM) Higgs boson~\cite{sm}.
The most recent
global fit to the electroweak precision data yields an estimate of the
SM Higgs boson mass $m_H=91^{+45}_{-32}\ \textrm{GeV}/c^{2}$ or $m_H
\le 186\,\textrm{GeV}/c^{2}$ at 95\% confidence level
(C.L.)~\cite{newewfit}.  This estimate is consistent with results from
direct searches at LEP that place a limit on the SM Higgs mass $m_H\ge
114\,\textrm{GeV}/c^{2}$ at 95\% C.L.~\cite{lep}.  Higgs boson
searches at the Tevatron are expected to be sensitive to Higgs
production at masses near the LEP limit~\cite{sensitivity}.  The CDF
collaboration has performed searches for new particles that decay into
$b\bar{b}$ using $p\bar p$ collision data with
$\sqrt{s}=1.8\,\mathrm{TeV}$~\cite{run1}.  Similar searches have been
pursued by the D\O\ collaboration using $p\bar p$ collision data at
$1.96\,\mathrm{TeV}$~\cite{d0}.

In this Letter we present an updated search for Higgs bosons and other
particles decaying into $b\bar{b}$ and produced in association with a
$W$ boson using data from Run II of the Tevatron collected with the
Collider Detector at Fermilab (CDF II).  The data sample corresponds
to an integrated luminosity of $320\,\textrm{pb}^{-1}$, nearly three
times that collected during Run I. In addition, we expect a 20\%
increase of the $WH$ production cross section due to the increase of
the center-of-mass energy from $1.8\,\textrm{TeV}$ (Run I) to
$1.96\,\textrm{TeV}$ (Run II). The standard model production cross
section for $m_H=115\,\textrm{GeV}/c^2$ is predicted to be
$0.2\,\textrm{pb}$ at the higher energy~\cite{smhiggsxsec}.  The
branching fraction $H\rightarrow b\bar b$ for that mass is
0.73~\cite{hdecay}.

The upgrades to the CDF detector increase the signal yield with
improved lepton acceptance and $b$ quark identification.  The search
signature considered here is $WH$ with $W \rightarrow e\nu$ or
$\mu\nu$ and with $H \rightarrow b\bar{b}$, giving final states with
one high-$p_{T}$ lepton, large missing $E_{T}$ ($\met$), and two
$b$-quark jets~\cite{eta}.  Requiring an associated $W$ boson reduces
the contribution from other physics processes.  We focus our attention
on the $W+2~\textrm{jets}$ signature with at least one identifed $b$
jet since that signature contains most of the signal, while $b$-tagged
$W+\ge3~\textrm{jets}$ events are dominated by $t \bar{t}$ production.
 
The CDF II detector~\cite{CDF} is an azimuthally and
forward-backward symmetric apparatus designed to study $p\bar p$
collisions at the Tevatron. The detector has a charged particle
tracking system immersed in a $1.4\,\textrm{T}$ magnetic field aligned
coaxially with the colliding beams. A silicon microstrip detector
provides tracking over the radial range from 1.5 to $28\,\textrm{cm}$.
A $3.1\,\textrm{m}$ long open-cell drift chamber, the Central Outer
Tracker (COT), covers the radial range from 40 to $137\,\textrm{cm}$.
The fiducial region of the silicon detector extends to absolute
pseudorapidity $|\eta|\approx 2$, while the COT provides coverage for
$|\eta|\le 1$.
 
Segmented electromagnetic and hadronic sampling calorimeters surround
the tracking system and measure particle energies in the
pseudo-rapidity range $|\eta|<3.6$.  A set of drift chambers located
outside the central hadron calorimeters and another set behind a
$60\,\textrm{cm}$ iron shield detect tracks from muon candidates with
$|\eta|\le 0.6$.  Additional drift chambers and scintillation counters
detect muons in the region $0.6\le |\eta |\le 1.0$.  The beam
luminosity is determined using gas Cherenkov counters located in the
$3.7<|\eta|<4.7$ region; the counters measure the average number of
inelastic $p\bar p$ collisions per bunch crossing.  The data used in
this analysis are collected with high-$E_T$ electron triggers or
high-$p_T$ muon triggers.

The signature of 2 jets plus a leptonically-decaying $W$ boson
allows us to make stringent event selection requirements. 
The event selection begins with the requirement of a primary lepton,
either an isolated electron with $E_{T} > 20\,\textrm{GeV}$ or an
isolated muon with $p_{T}> 20\,\textrm{GeV}/c$, in the central region
($|\eta|<1.0$).  The lepton isolation ($I_\textrm{cal}$) is defined by
the additional energy deposited in a calorimeter cone of radius
$\Delta R = \sqrt {(\Delta \phi)^2+(\Delta \eta)^2}=0.4$ around the
lepton. We require $I_\textrm{cal}$ be less than 10\% of the lepton
momentum.  A $W$ boson sample is selected by additionally requiring
$\met > 20\,\textrm{GeV}$.  Events that contain a second same-flavor
opposite-charge lepton with $E_T (p_T)$ $>$ 10 GeV (10 GeV/$c$) are removed as
possible $Z$ boson candidates if the reconstructed dilepton
invariant mass is between 76 and 106 GeV/$c^{2}$; these events are
considered by a parallel search at CDF in the $ZH$ production channel.
We also reject cosmic ray muon events and dilepton events identified
as top pair production using the selection in Ref.~\cite{topdil}.  To
further reduce this top dilepton background, we reject events with an
additional high-$p_{T}$ isolated track ($p_{T}$ $>$ 20 GeV/$c$) whose
charge is opposite that of the primary lepton.  The track isolation
requirement is that the total $p_{T}$ of all tracks within a cone of
radius $\Delta R = \sqrt {(\Delta \phi)^2+(\Delta \eta)^2}=0.4$
around, but excluding, the candidate track, be less than 2.0 GeV/$c$.
The remaining events are classified according to jet multiplicity.  A
jet is defined as a cluster of energy deposited in the calorimeter
within a fixed radius $\Delta R=0.4$.  Candidate events are required
to have two jets each with 
$E_{T}>15\,\textrm{GeV}$ and $|\eta|<2.0$ after corrections for
calorimeter response and multiple interactions.  To remove other
contributions from $t\bar t$ production, we veto events that have
additional jets with $E_T > 8\,\textrm{GeV}$ and $2.0 < |\eta| < 3.6$
as well as events that have jets with $8 < E_T < 15\,\textrm{GeV}$ and
$|\eta| < 2.0$.

The sample of events with $W+2~\textrm{jets}$ is expected to
contain most of the signal, while the sample of events with only 1 jet
is used to constrain the $W$ + heavy flavor ($b\bar b$ and
$c\bar c$) background, and the sample of events with 3 or more jets
is used to estimate the $t\bar t$ contribution.
In order to enhance the signal purity of the $W+2~\textrm{jets}$
sample, we require at least one jet be tagged by a secondary vertex
algorithm (SECVTX)~\cite{toplj} as containing a $B$ hadron.  
The SECVTX $b$ tagging algorithm searches for
secondary vertices formed by two or more tracks displaced with respect
to the primary event vertex.  A jet is
declared tagged if it contains a secondary vertex with a significant
transverse displacement from the primary vertex, positive if the
displacement is along the jet direction and negative if opposite the
jet direction.

Background events come predominantly from the direct production of $W$
bosons in association with heavy quarks ($Wb\bar{b}$, $Wc\bar{c}$,
$Wc$), $W$ production in association with light flavor jets that are
falsely tagged as $b$-quark jets, multi-jet events without $W$ bosons,
and $t\bar{t}$ and single top production.  Other relatively small
backgrounds include diboson ($W^{+}W^{-}$, $ZZ$, or $WZ$) and $Z
\rightarrow \tau^{+}\tau^{-}$ production.

The fraction of $W+\textrm{jets}$ events that contain heavy
quarks is estimated using the {\sc alpgen}~\cite{ALPGEN} Monte Carlo
program and is further calibrated using the inclusive jet
data~\cite{toplj}.  We find that this fraction must be scaled by an
additional factor of $1.2\pm0.2$ to match the observed number of
$b$ tagged $W+1~\textrm{jet}$ events.  The estimated number of
background events is then obtained by multiplying the heavy flavor
fraction, the event $b$ tagging efficiency, and the number of
$W+\textrm{jets}$ events in the data before $b$ tagging. The largest
uncertainty on this estimate comes from varying the parton shower
cutoff scale and the initial and final state radiation effects.

To determine the number of false $b$ tags of light-flavor jets, we
first parameterize the negative tag rate in an inclusive jet
sample~\cite{toplj} as a function of jet $E_{T}$, track multiplicity,
$\eta$, $\phi$, and the summed $E_{T}$ of all the jets in the event.
Negative tags are a good approximation of false positive tags, since
false tags are mostly due to symmetric resolution effects.  Long-lived
particles and secondary interactions with the detector material
contribute asymmetrically to the positive tag rate and cannot be
predicted from the negative tags.  By analyzing inclusive jet data, we
find the rate of negative tags must be increased by a factor of
$1.3\pm 0.1$ to account for these asymmetries. We apply the same
correction to the $W+\textrm{jets}$ sample to estimate the expected
number of false positive tagged events.

The remaining backgrounds are estimated from a combination of Monte
Carlo simulation and data.  The diboson and $Z\rightarrow
\tau^{+}\tau^{-}$ contributions are estimated from {\sc pythia} Monte
Carlo events.  A substantial background contribution comes from events
without $W$ bosons.  In these events a jet is misidentified as a
high-$p_{T}$ lepton, and mismeasured energies produce apparent missing
$E_{T}$. We measure this ``non-$W$'' background by extrapolating the
number of tagged events with an isolated lepton and low $\met$ into
the signal region~\cite{toplj}.  The top quark production
contributions are estimated using {\sc herwig}~\cite{herwig} and {\sc
  madevent+pythia}~\cite{madevent,pythia} Monte Carlo calculations for
$t\bar t$ and single top, respectively; $t\bar t$ production is
normalized to the cross section measured in this dataset
($\sigma_{t\bar t}=8.6\pm 1.3\,\textrm{pb}$)~\cite{salsthesis}, while
the single top estimate uses cross section
calculations~\cite{singletop} for a top quark mass of
$175\,\textrm{GeV}/c^{2}$.
 
The number of observed $b$ tagged events and the 
corresponding background estimates are listed in Table~\ref{tab:event_summary}. 
We find good agreement between data and background expectations. 
  
\begin{table} 
  \begin{center} 
    \caption{The number of observed $W+2~\textrm{jet}$ events with
      $\ge$~1 and $\ge$~2 tagged jets and the background summary for
      an integrated luminosity of $320$ pb$^{-1}$.} 
    \begin{tabular}{ccc} \hline\hline
      Number of tags:                         & $\ge$ 1 tags   & $\ge$ 2 tags                 \\ \hline 
      False tags                                      & $39.3 \pm 3.1$            & $ 1.0 \pm 0.1$    \\ 
      $Wb\bar{b}$                            & $54.0 \pm 18.4$           & $ 8.0 \pm 3.0$    \\ 
      $Wc\bar{c}$                            & $19.5 \pm 6.6$            & $ 0.4 \pm 0.2$    \\ 
      $Wc$                                   & $16.8 \pm 4.3$            & $ 0.0^{+0.1}_{-0.1}$    \\ 
      Diboson/$Z \rightarrow \tau^{+}\tau^{-}$ & $ 5.0 \pm 1.1$            & $ 0.3 \pm 0.1$    \\ 
      non-$W$                                      & $16.5 \pm 3.2$            & $ 0.4 \pm 0.1$    \\ 
      Single top                                   & $ 9.6 \pm 2.0$            & $ 1.3 \pm 0.3$    \\ 
      $t\bar{t}$                                   & $14.6 \pm 2.5$            & $ 3.1 \pm 0.5$    \\ \hline
      Total background                             & $175\pm 26$               & $15 \pm 3$        \\ \hline 
      Observed positive tagged events                       & $187$                     & $14$              \\ \hline\hline 
      Events before $b$ tagging                        & \multicolumn{2}{c}{$2072$}                    \\ \hline
    \end{tabular} 
    \label{tab:event_summary} 
  \end{center} 
\end{table} 
 
The acceptance for identifying $WH\rightarrow \ell \nu b\bar{b}$
events using the cuts described above is calculated as a function of
the Higgs boson mass from {\sc pythia} Monte Carlo samples of $WH
\rightarrow Wb\bar{b}$ events after full simulation of the CDF II
detector response.  The total acceptance is calculated as a product of
the $W$ leptonic branching fraction, the kinematic and geometric
acceptance, the lepton identification efficiencies, the trigger
efficiencies, and the $b$ tagging efficiencies. An $11\%$ systematic
uncertainty comes from uncertainties in the modeling of initial
($3\%$) and final ($7\%$) state radiation, parton distribution
function ($1\%$), jet energy scale ($3\%$), $b$ tagging efficiency
($5\%$), jet energy resolution (3\%), electron and muon trigger
efficiencies ($<1\%$), electron and muon ID efficiencies ($5\%$), and
integrated luminosity ($6\%$).  Acceptance for $WH$ increases linearly
from $1.5\%$ to $1.7\%$ as $m_{H}$ increases from $110$ to $150$
GeV$/c^{2}$.  For a Higgs boson mass of $115\,\textrm{GeV}/c^2$, we
expect 0.65 $WH$ events with an average efficiency of
$50\pm 3\%$ to tag at least one $b$ jet.
 
\begin{figure}[htbp] 
\begin{center} 
\includegraphics[width=0.45\textwidth,clip]{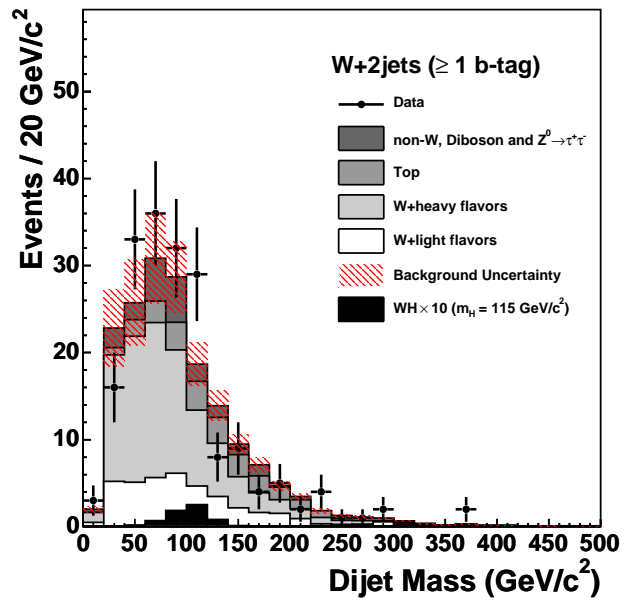} 
\caption{The dijet mass distribution in the data along with the background  
expectations and the Higgs boson signal distribution (for mass 115 GeV$/c^{2}$)
scaled up by a factor of 10.} 
\label{fig:dijet_mass} 
\end{center} 
\end{figure}

We perform a direct search for a resonant mass peak in the
reconstructed dijet invariant mass distribution, after having
corrected for the calorimeter response to $B$-hadron jets.  The
reconstructed $b\bar b$ invariant mass resolution for $WH$ events is
estimated to be 17\%.  Figure~\ref{fig:dijet_mass} shows the dijet
mass distribution in the $\ge1$-tag data.  The expected mass
distribution is also shown for the standard model Higgs boson but with
a cross section 10 times the expected value~\cite{smhiggsxsec}.  The
data match well the total predicted background, and we set an upper
limit on the production cross section times branching ratio of
$\sigma(p\bar{p} \rightarrow WH) \times \mathcal{B}(H \rightarrow
b\bar b)$ 
as a function of $m_H$
using the number of events in the $W+2~\textrm{jets}$ sample.
 
We assume the data consist of three components: TOP ($t\bar{t}$ and
single top), $WH$, and other SM processes labelled OTH (false tags,
$Wb\bar{b}$, $Wc\bar{c}$, $Wc$, non-$W$, and
diboson).  A binned maximum likelihood technique is used to estimate
upper limits on $WH$ production by constraining the number of
OTH and TOP events to the background expectations within their
separate uncertainties.  The expected number of events $\mu$ in each
mass bin is
    \[ 
    \mu = f_{\textrm{OTH}} \cdot N_{\textrm{OTH}} + f_{\textrm{TOP}}
    \cdot N_{\textrm{TOP}} + f_{WH} \cdot (\varepsilon \cdot {\cal{L}}
    \cdot \sigma_{WH} \cdot \mathcal{B}(H \rightarrow b\bar{b})),
    \] 
    where $f_{\textrm{OTH}}$, $f_{\textrm{TOP}}$ and $f_{WH}$ are the
    expected fraction of events in a given mass bin predicted by Monte
    Carlo simulation and $N_{\textrm{TOP}}$, $N_{\textrm{OTH}}$, $\varepsilon$,
    $\cal{L}$ and $\sigma_{WH}$ are the expected number of top and
    non-top events, the detection efficiency, the integrated luminosity and the
    $WH$ cross section to be estimated, respectively.  The corresponding
    likelihood is
    \[ 
    L = \prod_{i=\textrm{bin}} \frac{\mu_{i}^{N_{i}} \cdot
      e^{-\mu_{i}}}{N_{i}!} \times
    G(N_{\textrm{OTH}},\sigma_{\textrm{OTH}}) \times
    G(N_{\textrm{TOP}}, \sigma_{\textrm{TOP}}),
    \] 
    where $N_{i}$ is the observed events from
    $W+2~\textrm{jets}$ sample and $G$ is a Gaussian constraint,
    with width equal to the respective uncertainties,
    on the estimate of top and other background events from the
    counting experiment.
 
\begin{figure}[htbp] 
\begin{center} 
\includegraphics[width=0.45\textwidth,clip]{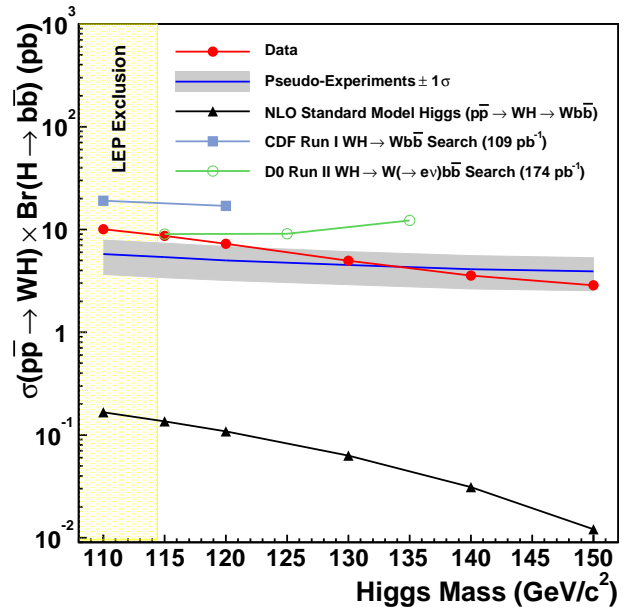} 
\caption{The $95\%$ C.L. upper limits (filled circles) on the $WH$ cross section 
  as a function of the Higgs boson mass for events with at least one $b$
  tag. Also shown is the theoretical
  cross section (triangles) for the production of a standard model
  Higgs boson in association with a $W$ boson, and the expected
  results from background-only experiments (shaded band).  The CDF
  Run I limits and the D\O\ Run II limits are also shown for
  comparison.}
\label{fig:wx_lvbb_limit} 
\end{center} 
\end{figure}

    The resulting $95\%$ confidence level upper limits range from 10
    to 3 pb as a function of the Higgs mass and are shown in
    Figure~\ref{fig:wx_lvbb_limit} in comparison with the production
    cross section times branching fraction for the standard model
    Higgs boson.  We generate background-only pseudo-experiments to
    calculate the expected performance of the analysis in the absence
    of signal events; the median results are also shown in
    Figure~\ref{fig:wx_lvbb_limit}.  These data are consistent with
    our expectations for a data sample of this size.
 
    We have studied additional systematic effects related to the dijet
    mass distribution, including uncertainties in the $b$-jet energy
    measurement and the $Wb\bar{b}$ and $t\bar{t}$ two-jet invariant
    mass spectra; these have a negligible effect on the calculated
    upper limits, as do the correlations between the uncertainties for
    different backgrounds.
 
\begin{figure}[htbp] 
\begin{center} 
\includegraphics[width=0.45\textwidth,clip]{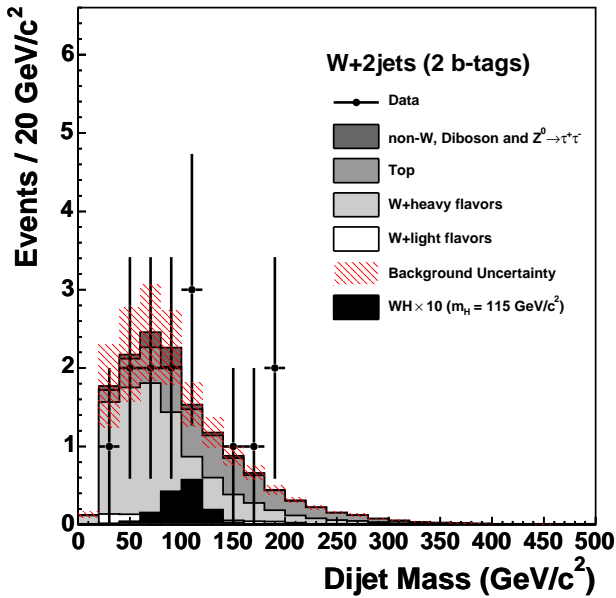} 
\caption{The dijet mass distribution in the double $b$-tagged data along with the background  
expectations and the Higgs boson mass signal distribution (for mass
115 GeV$/c^{2}$) scaled up by a factor of 10.} 
\label{fig:dijet_mass_double} 
\end{center} 
\end{figure}

A closer look at the sample of double $b$-tagged events, with both
jets tagged by SECVTX, still shows good agreement between the expected
backgrounds and the observed number of data events
(Table~\ref{tab:event_summary}).  The corresponding dijet invariant
mass distributions are shown in Figure~\ref{fig:dijet_mass_double}.  The total
acceptance is reduced significantly with respect to the single-tagged
sample, ranging from $0.32\pm 0.05\% $ to $0.40\pm 0.07\% $ for
$m_{H}$ between 110 and 150 GeV/c$^2$, where the errors include both
statistical and systematic uncertainties described above.  This
reduced acceptance hampers the sensitivity of the double-tagged
analysis with the current dataset; the sensitivity figure of merit
$S/\sqrt{B}$ drops from 0.07 for the single-tagged sample to 0.05 for
the double-tagged sample, assuming a Higgs mass of
$115\,\textrm{GeV}/c^2$.  We do not treat the double-tagged events as
a separate search dataset because the event selection has not been
optimized for that purpose.

In summary, this search with the CDF II detector is sensitive to new particles
decaying into $b\bar{b}$ and produced in association with a $W$
boson. The dijet mass spectrum
shows no significant excess of events in the $b$-tagged
$W+2~\textrm{jets}$ events. With a $320\,\textrm{pb}^{-1}$
dataset, we set $95\%$ C.L. upper limits on
the production cross section times branching ratio ranging from
$10\,\textrm{pb}$ at $m_H=110\,\textrm{GeV}/c^2$ to $3\,\textrm{pb}$
at $m_H=150\,\textrm{GeV}/c^2$.
Although we have significantly improved the upper
limits over Run I~\cite{run1}, the sensitivity of the present search
is limited by statistics to a cross section approximately a factor of 50
higher than the predicted cross section for standard model
Higgs boson production.  Improvements in the $b$-tagging efficiency,
lepton identification, and dijet mass resolution promise increased
sensitivity for future searches with increased integrated luminosity.

We thank the Fermilab staff and the technical staffs of the
participating institutions for their vital contributions. This work
was supported by the U.S. Department of Energy and National Science
Foundation; the Italian Istituto Nazionale di Fisica Nucleare; the
Ministry of Education, Culture, Sports, Science and Technology of
Japan; the Natural Sciences and Engineering Research Council of
Canada; the National Science Council of the Republic of China; the
Swiss National Science Foundation; the A.P. Sloan Foundation; the
Bundesministerium f\"ur Bildung und Forschung, Germany; the Korean
Science and Engineering Foundation and the Korean Research Foundation;
the Particle Physics and Astronomy Research Council and the Royal
Society, UK; the Russian Foundation for Basic Research; the Comisi\'on
Interministerial de Ciencia y Tecnolog\'{\i}a, Spain; in part by the
European Community's Human Potential Programme under contract
HPRN-CT-2002-00292; and the Academy of Finland.

\end{document}

%% file: July_2005_Authors.tex
\affiliation{Institute of Physics, Academia Sinica, Taipei, Taiwan 11529, Republic of China} 
\affiliation{Argonne National Laboratory, Argonne, Illinois 60439} 
\affiliation{Institut de Fisica d'Altes Energies, Universitat Autonoma de Barcelona, E-08193, Bellaterra (Barcelona), Spain} 
\affiliation{Baylor University, Waco, Texas  76798} 
\affiliation{Istituto Nazionale di Fisica Nucleare, University of Bologna, I-40127 Bologna, Italy} 
\affiliation{Brandeis University, Waltham, Massachusetts 02254} 
\affiliation{University of California, Davis, Davis, California  95616} 
\affiliation{University of California, Los Angeles, Los Angeles, California  90024} 
\affiliation{University of California, San Diego, La Jolla, California  92093} 
\affiliation{University of California, Santa Barbara, Santa Barbara, California 93106} 
\affiliation{Instituto de Fisica de Cantabria, CSIC-University of Cantabria, 39005 Santander, Spain} 
\affiliation{Carnegie Mellon University, Pittsburgh, PA  15213} 
\affiliation{Enrico Fermi Institute, University of Chicago, Chicago, Illinois 60637} 
\affiliation{Joint Institute for Nuclear Research, RU-141980 Dubna, Russia} 
\affiliation{Duke University, Durham, North Carolina  27708} 
\affiliation{Fermi National Accelerator Laboratory, Batavia, Illinois 60510} 
\affiliation{University of Florida, Gainesville, Florida  32611} 
\affiliation{Laboratori Nazionali di Frascati, Istituto Nazionale di Fisica Nucleare, I-00044 Frascati, Italy} 
\affiliation{University of Geneva, CH-1211 Geneva 4, Switzerland} 
\affiliation{Glasgow University, Glasgow G12 8QQ, United Kingdom} 
\affiliation{Harvard University, Cambridge, Massachusetts 02138} 
\affiliation{Division of High Energy Physics, Department of Physics, University of Helsinki and Helsinki Institute of Physics, FIN-00014, Helsinki, Finland} 
\affiliation{University of Illinois, Urbana, Illinois 61801} 
\affiliation{The Johns Hopkins University, Baltimore, Maryland 21218} 
\affiliation{Institut f\"{u}r Experimentelle Kernphysik, Universit\"{a}t Karlsruhe, 76128 Karlsruhe, Germany} 
\affiliation{High Energy Accelerator Research Organization (KEK), Tsukuba, Ibaraki 305, Japan} 
\affiliation{Center for High Energy Physics: Kyungpook National University, Taegu 702-701; Seoul National University, Seoul 151-742; and SungKyunKwan University, Suwon 440-746; Korea} 
\affiliation{Ernest Orlando Lawrence Berkeley National Laboratory, Berkeley, California 94720} 
\affiliation{University of Liverpool, Liverpool L69 7ZE, United Kingdom} 
\affiliation{University College London, London WC1E 6BT, United Kingdom} 
\affiliation{Massachusetts Institute of Technology, Cambridge, Massachusetts  02139} 
\affiliation{Institute of Particle Physics: McGill University, Montr\'{e}al, Canada H3A~2T8; and University of Toronto, Toronto, Canada M5S~1A7} 
\affiliation{University of Michigan, Ann Arbor, Michigan 48109} 
\affiliation{Michigan State University, East Lansing, Michigan  48824} 
\affiliation{Institution for Theoretical and Experimental Physics, ITEP, Moscow 117259, Russia} 
\affiliation{University of New Mexico, Albuquerque, New Mexico 87131} 
\affiliation{Northwestern University, Evanston, Illinois  60208} 
\affiliation{The Ohio State University, Columbus, Ohio  43210} 
\affiliation{Okayama University, Okayama 700-8530, Japan} 
\affiliation{Osaka City University, Osaka 588, Japan} 
\affiliation{University of Oxford, Oxford OX1 3RH, United Kingdom} 
\affiliation{University of Padova, Istituto Nazionale di Fisica Nucleare, Sezione di Padova-Trento, I-35131 Padova, Italy} 
\affiliation{LPNHE-Universite de Paris 6/IN2P3-CNRS} 
\affiliation{University of Pennsylvania, Philadelphia, Pennsylvania 19104} 
\affiliation{Istituto Nazionale di Fisica Nucleare Pisa, Universities of Pisa, Siena and Scuola Normale Superiore, I-56127 Pisa, Italy} 
\affiliation{University of Pittsburgh, Pittsburgh, Pennsylvania 15260} 
\affiliation{Purdue University, West Lafayette, Indiana 47907} 
\affiliation{University of Rochester, Rochester, New York 14627} 
\affiliation{The Rockefeller University, New York, New York 10021} 
\affiliation{Istituto Nazionale di Fisica Nucleare, Sezione di Roma 1, University of Rome ``La Sapienza," I-00185 Roma, Italy} 
\affiliation{Rutgers University, Piscataway, New Jersey 08855} 
\affiliation{Texas A\&M University, College Station, Texas 77843} 
\affiliation{Istituto Nazionale di Fisica Nucleare, University of Trieste/\ Udine, Italy} 
\affiliation{University of Tsukuba, Tsukuba, Ibaraki 305, Japan} 
\affiliation{Tufts University, Medford, Massachusetts 02155} 
\affiliation{Waseda University, Tokyo 169, Japan} 
\affiliation{Wayne State University, Detroit, Michigan  48201} 
\affiliation{University of Wisconsin, Madison, Wisconsin 53706} 
\affiliation{Yale University, New Haven, Connecticut 06520} 
\author{A.~Abulencia}
\affiliation{University of Illinois, Urbana, Illinois 61801}
\author{D.~Acosta}
\affiliation{University of Florida, Gainesville, Florida  32611}
\author{J.~Adelman}
\affiliation{Enrico Fermi Institute, University of Chicago, Chicago, Illinois 60637}
\author{T.~Affolder}
\affiliation{University of California, Santa Barbara, Santa Barbara, California 93106}
\author{T.~Akimoto}
\affiliation{University of Tsukuba, Tsukuba, Ibaraki 305, Japan}
\author{M.G.~Albrow}
\affiliation{Fermi National Accelerator Laboratory, Batavia, Illinois 60510}
\author{D.~Ambrose}
\affiliation{Fermi National Accelerator Laboratory, Batavia, Illinois 60510}
\author{S.~Amerio}
\affiliation{University of Padova, Istituto Nazionale di Fisica Nucleare, Sezione di Padova-Trento, I-35131 Padova, Italy}
\author{D.~Amidei}
\affiliation{University of Michigan, Ann Arbor, Michigan 48109}
\author{A.~Anastassov}
\affiliation{Rutgers University, Piscataway, New Jersey 08855}
% ...
\collaboration{CDF Collaboration}
\noaffiliation

%% file: wh_prl.bbl
\begin{thebibliography}{99} 
    \bibitem{sm}  
    S. Weinberg, Phys. Rev. Lett. {\bf 19} (1967) 1264; A. Salam,
    {\it Proceedings of the 8th Nobel Symposium}, edited by
    N. Svartholm (Almqvist and Wiksell, Stockholm, 1968), p. 367;
    S. Glashow, Nucl. Phys.{\bf 22} (1961) 579.
    \bibitem{newewfit}
      LEP Electroweak Working Group, {\tt hep-ex/0511027}.
    \bibitem{lep}
    A. Heister {\it et al.}, (The LEP Higgs Working Group),
    Phys. Lett. B {\bf 565}, 61 (2003), {\tt hep-ex/0306033}.
    \bibitem{sensitivity} M. Carena {\it et al.}, Report of the Tevatron
    Higgs Working Group, hep-ph/0010338; CDF and D\O\ Collaborations,
    Results of the Tevatron Higgs Sensitivity Study,
    FERMILAB-PUB-03/320-E, {\tt hep-ph/0010338}.
    \bibitem{run1}
    F. Abe {\it et al.} (CDF Collaboration), Phys. Rev. Lett. {\bf
      79}, 3819 (1997);
    D. Acosta {\it et al.} (CDF Collaboration), Phys. Rev. Lett. {\bf
      95}, 051801 (2005), {\tt hep-ex/0503039}.
    \bibitem{d0} 
    V.M. Abazov {\it et al.} (D\O\ Collaboration),
    Phys. Rev. Lett. {\bf 94}, 091802 (2005), {\tt hep-ex/0410062}. 
    \bibitem{smhiggsxsec}
      T. Han and S. Willenbrock, Phys. Lett. B273 (1991) 167.
    \bibitem{hdecay}
      A. Djouadi, J. Kalinowski, and M. Spira, Comp. Phys. Commun. 108
      C (1998) 56, {\tt hep-ph/9704448}.
    \bibitem{eta}
      We use a cylindrical coordinate system about the beam axis in
      which $\theta$ is the polar angle, $\phi$ is the azimuthal
      angle, and pseudorapidity is defined $\eta = -\ln \tan
      (\theta/2)$.  We define $E_{T} = E\sin\theta$ and $p_{T} =
      p\sin\theta$ where $E$ is energy measured by the calorimeter and
      $p$ is momentum measured by the spectrometer. The missing
      transverse energy vector, $\met$, is $-\Sigma_{i} E^{i}_{T} \vec
      {n_i}$, where $\vec{n_i}$ is the unit vector in the azimuthal
      plane that points from the beamline to the $i$-th calorimeter
      tower.
    \bibitem{CDF}
     D. Acosta {\it et al.} (CDF Collaboration), Phys. Rev. D {\bf
       71}, 032001 (2005), {\tt hep-ex/0412071}.
    \bibitem{topdil}
     D. Acosta {\it et al.} (CDF Collaboration), Phys. Rev. Lett. {\bf
       93}, 142001 (2004), {\tt hep-ex/0404036}.
    \bibitem{toplj}
     D. Acosta {\it et al.} (CDF Collaboration), Phys. Rev. D. {\bf
       71}, 052003 (2005), {\tt hep-ex/0410041}.
    \bibitem{ALPGEN}
    M.L. Mangano {\it et al.}, JHEP {\bf 0307}, 001 (2003), {\tt hep-ph/0206293}.
    \bibitem{herwig}
    G. Corcella {\it et al.}, JHEP {\bf 0101}, 010 (2001), {\tt hep-ph/0011363}.
    \bibitem{madevent}
      F. Maltoni and T. Stelzer, JHEP {\bf 0302}, 027 (2003), {\tt hep-ph/0208156}.
    \bibitem{pythia}
    T. Sj\"ostrand {\it et al.}, Comput. Phys. Commun. {\bf 135}, 238 (2001), {\tt hep-ph/0010017}.
    \bibitem{salsthesis}
      S.R. Rappoccio, Ph.D. thesis, Harvard University, 2005.
    \bibitem{singletop}
    B.W. Harris and E. Laenen and L. Phaf and Z. Sullivan and
    S. Weinzierl, Phys. Rev. D {\bf 66}, 054024 (2002), {\tt hep-ph/0207055}.
\end{thebibliography}
